\documentclass[9pt, lettersize, journal]{IEEEtran}
\IEEEoverridecommandlockouts
\usepackage{graphicx}
\usepackage{color}
\usepackage[utf8]{inputenc}
\usepackage{amsmath,amssymb,amsfonts,amsthm}
\usepackage{algorithmic}
\usepackage{multirow}
\usepackage{stfloats}

\title{DeMM: A Decoupled Matrix Multiplication Engine Supporting Relaxed Structured Sparsity}

\author{Christodoulos Peltekis, Vasileios Titopoulos, Chrysostomos Nicopoulos and Giorgos Dimitrakopoulos
\thanks{This work was supported by a research grant from Codasip, a provider of customizable RISC-V IP and Codasip Studio design toolset, to DUTh.}
\thanks{Christodoulos Peltekis, Vasileios Titopoulos and Giorgos Dimitrakopoulos are with the Department of Electrical and Computer Engineering, Democritus University of Thrace, Xanthi, Greece,
(e-mail: cpeltekis@ee.duth.gr, vtitopou@ee.duth.gr, dimitrak@ee.duth.gr)}
\thanks{Chrysostomos Nicopoulos is with the Department of Electrical and Computer Engineering at the University of Cyprus, Nicosia, Cyprus (e-mail: nicopoulos@ucy.ac.cy).}
}

\begin{document}

\maketitle

\begin{abstract}
Deep Learning (DL) has achieved unprecedented success in various application domains. 
Meanwhile, model pruning has emerged as a viable solution to reduce the footprint of DL models in mobile applications, without compromising their accuracy. 
To enable the matrix engines built for dense DL models to also handle their pruned counterparts, pruned DL models follow a fine-grained structured sparsity pattern of 1:4, or 2:4, whereby in each group of four contiguous values, at least one, or two, respectively, must be non-zero. Structured sparsity has recently also moved to coarser (relaxed) cases of $N$:128, or $N$:256, for small values of $N$, targeting a wider range of sparsity (10\%-90\%) for the DL models. In this work, we design an accelerator that operates, by construction, on wide blocks with relaxed structured sparsity.
In contrast to the conventional systolic array archetype, the new engine decouples the memory part of the systolic array from the multiply-add units. The memory block comprises 1 write and $N$ read ports, with the number of read ports being equal to the number of non-zero elements per row. The multiply-add units connect directly to each read port and complete the multiplication in a row-wise product-first order. More importantly, simple reconfiguration facilitates more dense patterns.
The experimental evaluation demonstrates substantial latency improvements over current state-of-the-art systolic array engines built for fine-grained and relaxed structured sparsity.

\end{abstract}


\section{Introduction}
The acceleration of DL models, for both training and inference, relies primarily on equivalent matrix multiplications that inherently map to systolic arrays. To reduce memory storage and computation cost, the weights of DL models are pruned, thereby leading to sparse models~\cite{deep-compression, hoefler2021sparsity}. The derived zero weights are not stored and the corresponding computation is skipped. When sparsification occurs during training, the possible accuracy loss is ameliorated by allowing the model to adapt to the removal of certain weights. 

\begin{figure}[t!]
\centering
\includegraphics[width=0.84\columnwidth]{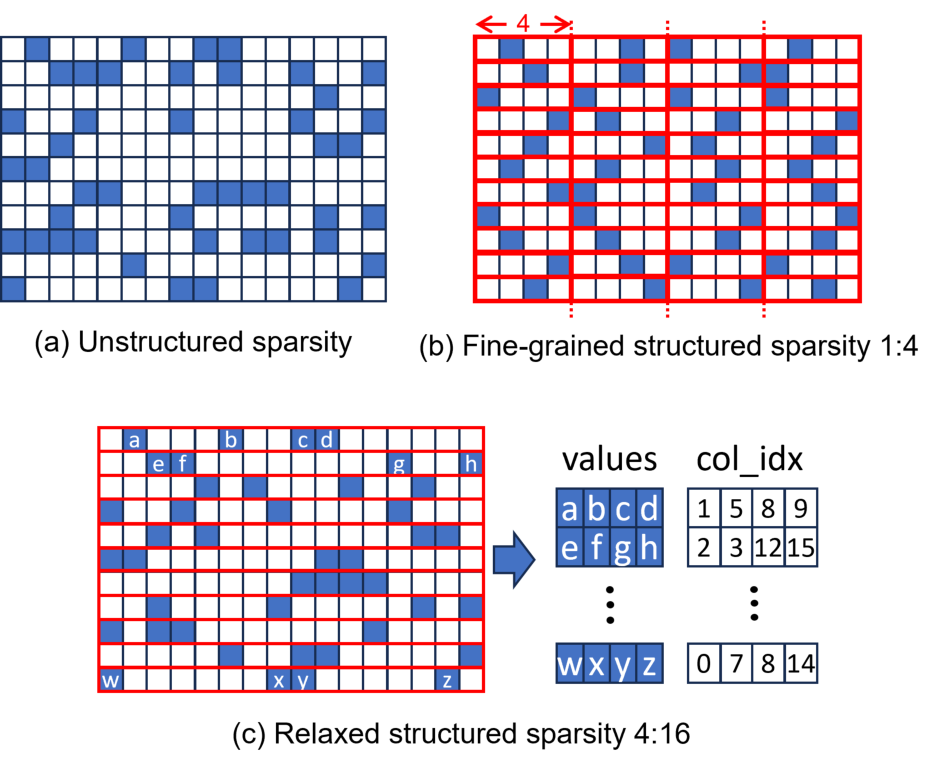}
\caption{Examples of (a) unstructured sparsity; (b) structured block sparsity of 1:4 (i.e., up to 1 non-zero element in every 4 consecutive elements); and (c) relaxed structured sparsity 4:16, and the corresponding packed representation of the non-zero elements. A blue square indicates a non-zero element.}
\label{f:unstructered-block-sparse}
\end{figure}

The achieved sparsity can either be \textit{unstructured}~\cite{rigl}, or \textit{structured}~\cite{nvidia-block-sparse,learning-n-m}. In unstructured sparsity, there is no constraint on the locations of the zeros, as shown in Fig.~\ref{f:unstructered-block-sparse}(a). In this case, together with the non-zero elements, multiple metadata indexes are also required to identify the original position of each non-zero element.

On the contrary, in structured sparsity, there is an upper limit on the number of non-zero elements that may be present within a block of consecutive elements (other forms of structured sparsity are also possible). For instance, in Fig.~\ref{f:unstructered-block-sparse}(b), for every 4 elements in each row, there is up to one non-zero element. Such structured sparsity simplifies both the indexing required to identify the position of each non-zero element inside each block, and the hardware needed to operate on such sparse data. This simplicity is of paramount importance to \textit{lightweight} engines (the focus of this work) found in mobile and embedded applications.

In most practical applications~\cite{nvidia-block-sparse,sta,s2ta}, blocks are small and \emph{fine-grained} $N$:$M$ sparsity patterns of 1:2, 1:4 or 2:4 are supported, where each block of $M$ elements may contain up to $N$ non-zero elements. Nevertheless, while fine-grained structured sparsity promises high performance and low storage overhead, it may also lead to less accurate ML models~\cite{hoefler2021sparsity,sparse-tensor-core}. This possible weakness is attributed to the constraints imposed during the fine-grained sparsification, where a fixed amount of non-zero elements is required for all consecutive small blocks.

To increase the flexibility during model training, sparsity could refer to much coarser blocks~\cite{vegeta, muralidharan2023uniform}. For instance, a \emph{relaxed} (coarser) structured sparsity of 8:128 allows the presence of at most 8 non-zero elements in every 128 consecutive elements. Fig.~\ref{f:unstructered-block-sparse}(c) shows an example of 4:16 relaxed structured sparsity, together with the packed representation of each row, which contains the non-zero element values and their corresponding column indexes.
Moving to coarser blocks complicates the operation of the corresponding hardware modules -- e.g., systolic arrays -- that operate optimally on well-structured data with small block sizes. 

To effectively reconcile these two conflicting attributes of relaxed (coarser) sparsity vs. hardware complexity, this work proposes a novel matrix-multiplication engine that supports \textit{relaxed} structured sparsity patterns, while still employing a simple and \emph{decoupled} hardware organization. 
Unlike conventional systolic arrays that co-locate the Multiply-Accumulate (MAC) and storage units within each tile, the proposed \underline{De}coupled \underline{M}atrix-\underline{M}ultiplication (DeMM) engine decouples the two. It essentially re-organizes the (dispersed) memory portion of a traditional systolic array into a regular standard-cell memory structure with multiple read ports. This transformation enables the support of relaxed structured sparsity and maintains the required regularities in the data flow and the physical layout.

Overall, the proposed DeMM engine provides a two-fold benefit. First, it enables support for relaxed structured sparsity patterns that combine hardware simplicity (similar to handling fine-grained sparsity) with additional flexibility during DL model pruning~\cite{hoefler2021sparsity,sparse-tensor-core}. Secondly, through appropriate reconfiguration, the DeMM engine can also support denser sparsity, which allows for the tackling of more common fine-grained structured sparsity patterns~\cite{nvidia-block-sparse}.

The experimental results demonstrate substantial improvements in overall execution latency over state-of-the-art matrix engines built to support fine-grained~\cite{s2ta} and relaxed structured sparsity~\cite{vegeta, spots} when executing structured-sparse CNN models. It should be noted that, even though said approaches -- including the proposed DeMM engine -- are effective for the low-sparsity levels of DNNs (i.e., 10\%-90\%), they are not as efficient in high-sparsity levels of above 95\%. At such high-sparsity levels, other accelerator architectures perform better. Examples include architectures following a non-systolic dataflow-like organization~\cite{sparse-abstract-machine, gamma} and ones that optimize memory traffic by operating either near-memory~\cite{spade, menda}, or in-memory~\cite{rram-dnn}.

\section{Simplifying Sparse$\times$Dense Matrix Multiplication}

The proposed DeMM engine employs a row-wise approach~\cite{matraptor} 
in computing the matrix product $A\times B$. Matrix $A$ follows a relaxed structured sparsity template and
$B$ is dense. 
The product of the multiplication is produced row-by-row, as follows: 
\begin{equation}
C[i,:] = \sum_{k} A[i,k] B[k,:]
\end{equation}
All the non-zero elements in a single row of matrix $A$ should be multiplied in parallel with the corresponding rows of matrix $B$, where the row index of matrix $B$ is determined by the column index of the non-zero value in matrix $A$.

\subsection{The proposed DeMM engine}

To achieve the desired parallelism, we \textit{decouple} the \textit{storage} (as used in a systolic array) from the \textit{multiply-add} units and treat each portion separately. Matrix $B$ is assumed to be pre-loaded in the storage area of DeMM. The pre-loading resembles the pre-loading operation of the input- (or weight-) stationary dataflow applied in systolic arrays. In each cycle, another row of $B$ is written into the memory block, using the one available write port. Subsequently, multiplication is computed row-by-row, by feeding the engine with all the non-zero elements of each row of stuctured-sparse matrix $A$.

\begin{figure}[t]
\centering
\includegraphics[width=0.62\columnwidth]{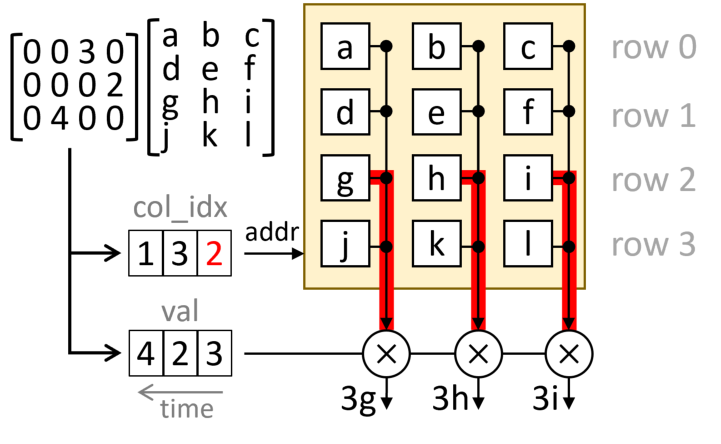}
\caption{Read and multiply operations are sufficient to perform matrix multiplication when the sparse matrix contains at most one non-zero element per row.}
\label{f:select-multiply}
\end{figure}

Let us initially assume the trivial case where each row of $A$ consists of at most one non-zero element. This element is passed to the engine, together with its column index, one after the other. For each \{value, column\_index\} pair, multiplication is performed in two serial steps, as depicted in Fig.~\ref{f:select-multiply}. First, the column index is treated as a row address to the memory that stores matrix $B$. This address allows us to read out all elements of the corresponding row of $B$. In the second step, the read elements are all multiplied in parallel with the value of the non-zero element of $A$. Repeating these two steps for all rows of matrix $A$ would complete the multiplication. The hardware engine required in this trivial case 
of a single non-zero element per row is just a memory block with 1 read port and a row of multipliers.

\begin{figure}
\centering
\includegraphics[width=0.67\columnwidth]{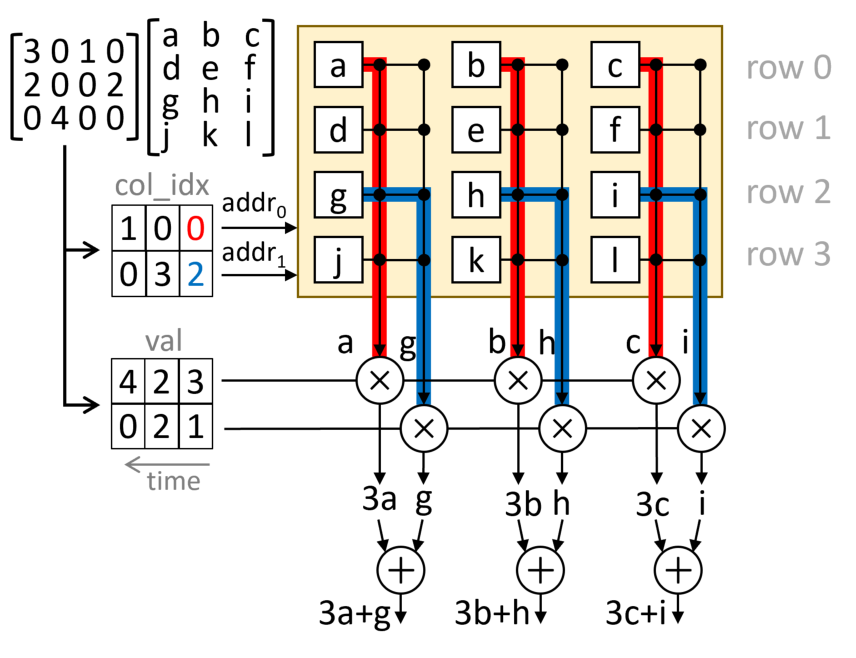}
\caption{Multiplying a sparse matrix with at most two non-zero elements per row requires two separate memory read ports and two rows of multipliers. The products of each port are then independently added in parallel to form the final result of the output row.}
\label{f:select-multiply-add}
\end{figure}

To support more than one non-zero element per row, one must simply increase the number of read ports in the memory block and, correspondingly, the number of multipliers per read port. Adders are also needed for the final addition. Fig.~\ref{f:select-multiply-add} illustrates an example of the operation of the proposed engine when operating with a row sparsity of two non-zero elements per row. In this case, the pairs of non-zero elements of each row of matrix $A$ are sent in parallel to the multiplication engine, one after the other. Each non-zero element is forwarded to its dedicated read port. The column index of each non-zero element selects (reads) the corresponding row of matrix $B$, and the value of each non-zero element is multiplied in parallel with all the elements of the selected row. The two products generated at each read port are added to finalize the result for this output row.

\begin{figure}
\centering
\includegraphics[width=0.82\columnwidth]{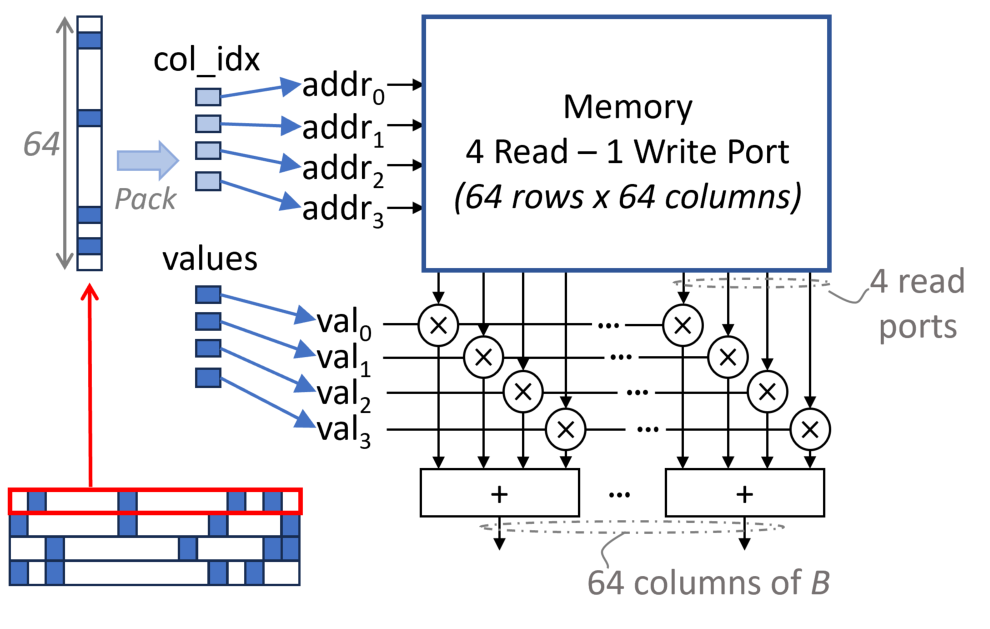}
\caption{The overall organization of a DeMM engine that supports relaxed structured sparsity of 4:64 using a memory block of four read ports and four multipliers and one add-reduction unit per output element. The example assumes that 64 outputs (columns) are computed in parallel.}
\label{f:overall}
\end{figure}

In the general case, matrix $A$ follows an $N$:$M$ row-sparsity pattern, where $M$ is much larger than $N$, e.g., $N$=8 and $M$=128, or $M$=256. Hence, the proposed DeMM engine consists of a regular memory of $N$ read and 1 write port. Each read port of DeMM's memory block outputs one data item per column, as selected by the column index address (`col\_idx') , which points to a row of matrix $B$. Fig.~\ref{f:overall} depicts a complete DeMM engine supporting 4:64 structured sparsity ($N$=4 and $M$=64).
Each read port is associated with a multiplier and the products of all read ports are reduced to one sum. Summation at the bottom of Fig.~\ref{f:overall} is implemented as a pipelined multi-operand adder of logarithmic depth.

\subsection{Supporting denser structured sparsity}

To support denser structured sparsity, e.g., $kN$:$M$, in a reconfigurable manner, DeMM should be able to read more than $N$ non-zero elements from \textit{the same} $M$ rows of matrix $B$. Since the memory block of each DeMM engine consists of $N$ read ports, it means that reading the $kN$ non-zero elements of the same row of $A$ requires time-sharing of the $N$ read ports for $k\times$ more cycles. To enable this sharing, every read port is associated with a $k$-to-$1$ multiplexer.
Note that, irrespective of the exact structured sparsity pattern supported, the memory of the DeMM engine is pre-loaded with the same $M$ rows of matrix $B$. The value of $k$ just determines how many times this block would be read before completing the computation for a row of $A$.

\begin{figure}
\centering
\includegraphics[width=0.85\columnwidth]{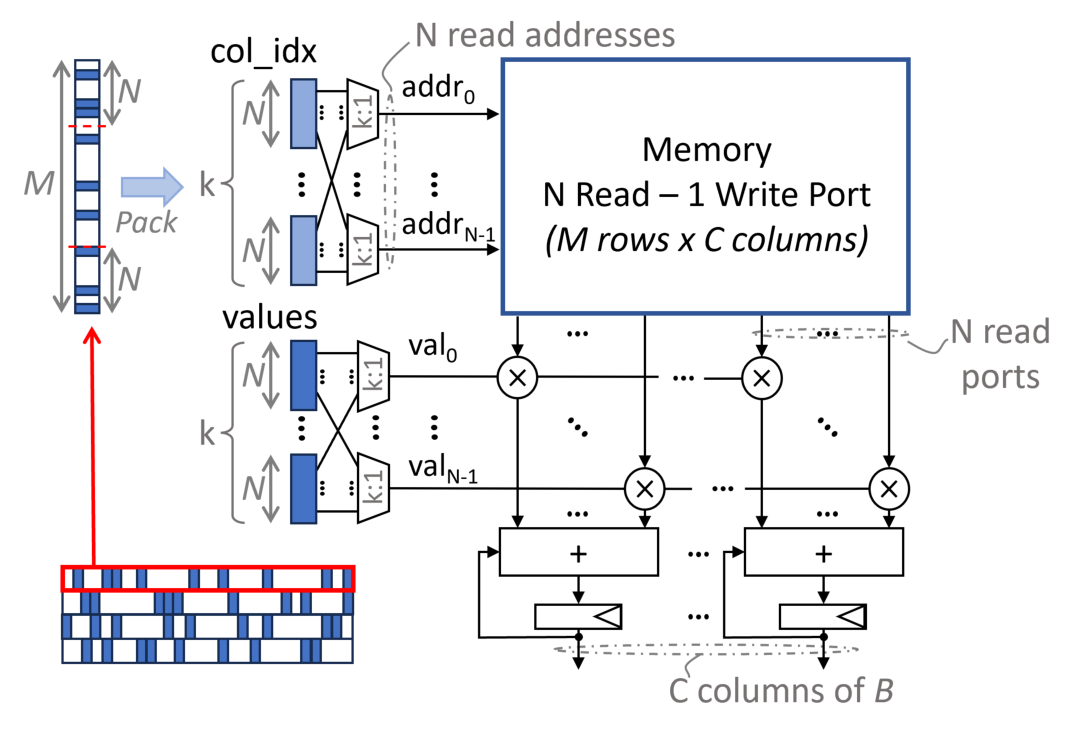}
\caption{The overall architecture of the DeMM engine that supports an $N$:$M$ relaxed structured sparsity and can be reconfigured for all $kN$:$M$ denser variants.}
\label{f:support-denser-format}
\end{figure}

The overall organization of the DeMM engine is depicted in Fig.~\ref{f:support-denser-format}.
The value chosen for $k$ reflects the reconfiguration properties of the proposed engine. For instance, by selecting $k$=4 for the  engine depicted in Fig.~\ref{f:overall}, which operates by default on 4:64 relaxed structured sparsity ($N$=4 and $M$=64), it means that all denser (and more fine-grained) structured sparsities can also be supported by the proposed design, e.g., 4:32 (as 8:64), 4:16 (as 16:64). Moving to an even denser scenario, such as 4:8, implemented in DeMM as a 32:64 pattern, would need larger multiplexers at each read port, i.e., $k$=8.

To identify the various design options, we define the DeMM($N, M, C, k$) configuration as the one that operates on a structured-sparse matrix $A$ of row size $M$ (this is also the number of rows in matrix $B$) and a matrix $B$ with $C$ columns. Structured sparsity can be as relaxed as $N$:$M$, or as dense as $kN$:$M$. 
The corresponding hardware block requires $N\times C$ multipliers, $C$ $N$-to-1 reduction trees, and an $M\times C$ memory block of $N$ read ports.

\begin{figure*}[b!]
    \centering
    \includegraphics[width=0.95\textwidth]{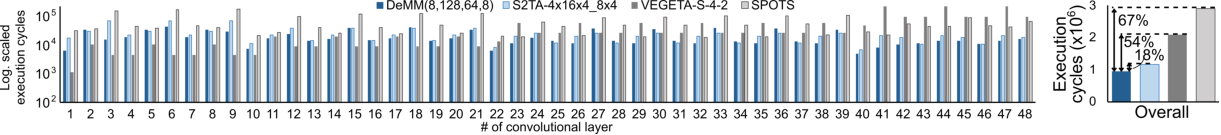}
    \caption{The execution latencies of all CNN layers of ResNet50~\cite{resnet}, when using the proposed DeMM(8,128, 64, 8) design, the S2TA-4$\times$16$\times$4\_8$\times$4 design~\cite{s2ta}, the VEGETA-S-4-2 design~\cite{vegeta}, and the SPOTS design~\cite{spots}. All three architectures have equal amount of computational resources.}
    \label{f:latency-1-16}
\end{figure*}

\section{Evaluation}

The effectiveness of the proposed decoupled matrix engine is evaluated by running inference in state-of-the-art CNN applications~\cite{resnet, convnext} with inputs from ImageNet. In the first set of experiments, we consider highly sparse CNN models derived from unstructured pruning. As reported in~\cite{hoefler2021sparsity, sparse-tensor-core}, unstructured pruning can achieve higher model compaction with better performance, as compared to structured pruning. In particular, we employ ResNet50~\cite{resnet}, pruned with RigL~\cite{rigl} at 95\% sparsity that roughly matches the relaxed sparsity of $8$:$128$ targeted by this work.
Any rows exceeding the sparsity of $8$:$128$ are computed in multiple consecutive cycles. For completeness, in the second set of experiments, we also include scenarios with fine-grained structured sparsity of $1$:$2$, $1$:$4$ and $1$:$8$, derived with Tensorflow for ResNet50~\cite{resnet} and ConvNeXt~\cite{convnext}.

\subsection{Relaxed structured sparsity}

For relaxed row sparsity of $8$:$128$, we compare DeMM with three state-of-the-art architectures: (a) VEGETA~\cite{vegeta}, which is able to support such sparsity degrees in a structured form of $1$:$16$ blocks, operating with a weight-stationary dataflow~\cite{scalesim}; (b) a version of S2TA~\cite{s2ta} configured to support block density $1$:$16$ using output stationarity; and (c) SPOTS~\cite{spots}, which skips groups of weights and input data that consist of only zero elements, following an output-stationary dataflow. Specifically, we compare DeMM(8,128,64,8) following an input-stationary dataflow to VEGETA-S-4-2~\cite{vegeta}, to S2TA-4$\times$16$\times$4\_8$\times$4~\cite{s2ta}, and to SPOTS~\cite{spots}, with \textit{all} designs under evaluation having \textit{the same amount of computational resources of 512 multiply-add units}. The evaluated VEGETA and S2TA designs have array sizes of $32\times 16$ PEs, while SPOTS has an array size of $128\times 4$ that can be reconfigured as four 32$\times$4 blocks operating in parallel. The configuration selected is the one that offers the best performance depending on the size of the input matrices.

The obtained results are summarized in Fig.~\ref{f:latency-1-16}, which shows the execution latencies of all CNN layers of ResNet50~\cite{resnet}.
DeMM's performance in the first layers is not the best, but it substantially outperforms the other three designs in the later layers. This behavior is the combined result of DeMM's engine architecture and the size of the stationary matrices in each case~\cite{scalesim}. DeMM   
leads to an \textit{overall} (across all CNN layers) {\bf latency improvement} of {\bf 18\%}, {\bf 54\%} and {\bf 67\%}, as compared to S2TA~\cite{s2ta}, VEGETA~\cite{vegeta}, and SPOTS~\cite{spots}, respectively.

\subsection{Hardware complexity}

All four evaluated designs were implemented in SystemVerilog and synthesized using the Cadence digital implementation flow and a 28 nm standard-cell library. The designs operate on 16-bit integer quantized inputs and weights, while the accumulations are performed with 32 bits. A block density of $8$:$128$ is used, which is DeMM's primary target. However, the instance of DeMM that is evaluated can also support more fine-grained patterns, down to the equivalent of 1:2. The equivalent density for S2TA and VEGETA is $1$:$16$. All designs under evaluation operate at a clock frequency of 500 MHz at 0.88 V.

\begin{figure}
    \centering
    \includegraphics[width=0.85\columnwidth]{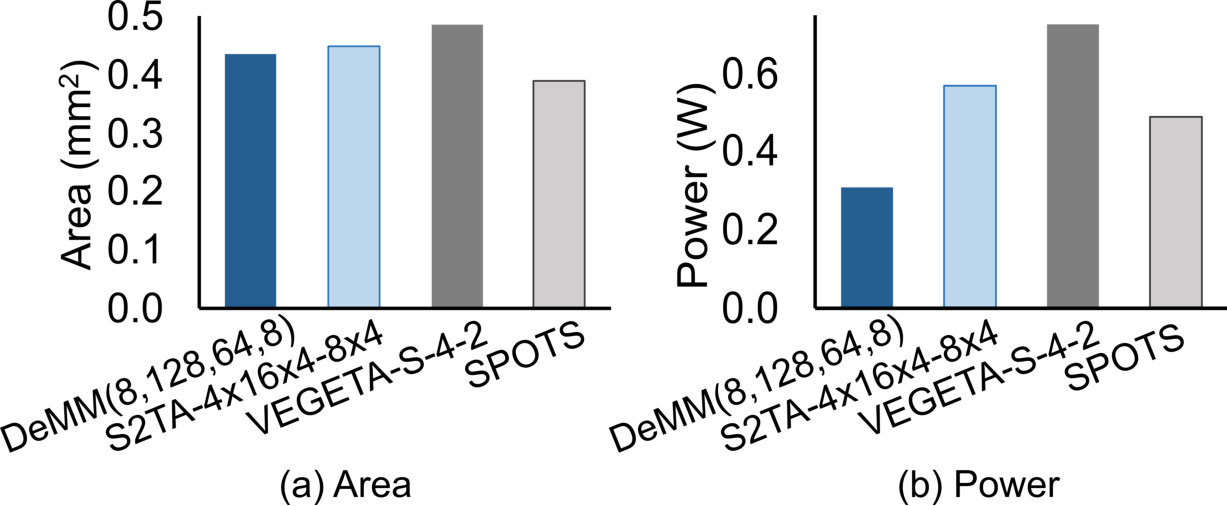}
    \caption{Hardware (a) area, and (b) power consumption comparisons between DeMM(8,128,64,8), S2TA-4$\times$16$\times$4\_8$\times$4~\cite{s2ta}, VEGETA-S-4-2~\cite{vegeta} and SPOTS~\cite{spots}. DeMM and SPOTS use a block density of $8$:$128$; S2TA and VEGETA use the equivalent $1$:$16$.}
    \label{f:area-power-1-16}
\end{figure}

Fig.~\ref{f:area-power-1-16}(a) compares the hardware area of the four architectures. Compared to S2TA and VEGETA, the DeMM engine requires {\bf 2.7\%} and {\bf 10.4\%}, respectively, {\bf lower area}, which is a testament to the simplicity of its organization. DeMM is slightly larger than SPOTS (the overhead is less than 10\%), due to the additional multiplexing logic required for supporting reconfigurability and multi-porting. Each additional read port added to the 128$\times$64 standard-cell-based memory block used in this DeMM setup costs 16\% more area.

In terms of power consumption, DeMM demonstrates significantly better behavior than the other designs. As shown in Fig.~\ref{f:area-power-1-16}(b), DeMM consumes {\bf 36.4\%}, {\bf 45.8\%} and {\bf 56.1\% lower power} than SPOTS, S2TA, and VEGETA, respectively. This substantial power reduction is mainly attributed to the minimization of data movement in pipeline registers. Both S2TA and VEGETA have an input data demand of a multiple-of-$M$ inputs and $C$ weights, while DeMM has a much lower input data demand of $C$ inputs and $kN$ weights, for a $kN$:$M$ structured sparsity, along with their addresses. The extra hardware cost of VEGETA relative to S2TA stems from its reconfiguration-rich organization that is also offered by the DeMM engine. SPOTS has a low multiplexing overhead that reduces its area requirements. However, its deeply pipelined operation increases its power consumption relative to DeMM.

\subsection{Fine-grained structured sparsity}

Furthermore, we also compare DeMM to VEGETA and S2TA in use-cases that better fit these two architectures, i.e., in scenarios with fine-grained block densities.
The designs are evaluated using ResNet50~\cite{resnet} and ConvNeXt~\cite{convnext}. The selected workloads are pruned to
fine-grained structured block sparsities of $1$:$2$, $1$:$4$ and $1$:$8$ to ensure optimal conditions for both VEGETA and S2TA, even if this choice is not the best option for DeMM, which inherently supports a wide range of sparsity formats. 
SPOTS is omitted in this comparison, since -- under such fine-grained structured sparsity -- it is very difficult to find contiguous groups of zero data, as required by SPOTS. Consequently, SPOTS exhibits significantly higher latencies.

\begin{figure}[h]
    \centering
    \includegraphics[width=0.95\columnwidth]{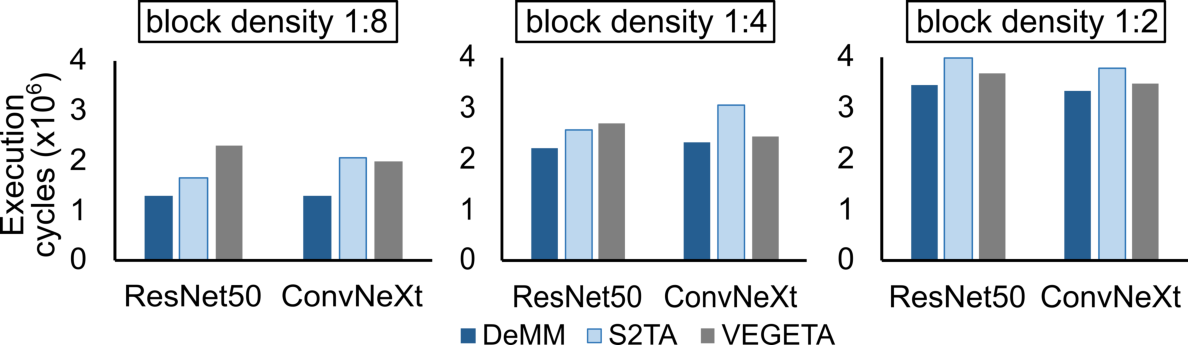} \\
    \caption{Overall execution latencies of ResNet50~\cite{resnet} and ConvNeXt~\cite{convnext} for configurations with block densities of $1$:$8$, $1$:$4$ and $1$:$2$.}
    \label{f:latency-1-8-4-2}
\end{figure}

The results are shown in Fig.~\ref{f:latency-1-8-4-2}. 
DeMM engine \textit{outperforms both} S2TA and VEGETA architectures in terms of overall latency. Specifically, for block sparsity $1$:$8$, DeMM achieves average {\bf latency improvements} of {\bf 29\%} and {\bf 39\%}, as compared to S2TA and VEGETA, respectively. For block sparsity $1$:$4$, DeMM's respective improvement in average latency is {\bf 19\%} and {\bf 12\%}. Finally, for block sparsity $1$:$2$, DeMM still yields {\bf 14\%} and {\bf 5\%} average latency improvements, as compared to S2TA and VEGETA, respectively.

\section{Conclusions}
This paper proposes DeMM, a matrix-multiplication engine that natively supports \textit{relaxed} structured sparsity patterns, without sacrificing the simplicity and regularity of the hardware organization. Contrary to conventional systolic arrays, the DeMM design employs a disaggregated micro-architecture that decouples the memory elements from the MAC units. Fine-grained sparsity patterns are also supported with minimal reconfiguration. The experimental evaluation demonstrates substantial improvements in the overall execution latency of modern CNN workloads, as compared to two existing state-of-the-art architectures~\cite{vegeta,s2ta, spots} for structured sparse workloads. Most importantly, these performance improvements are achieved with markedly lower power consumption.

\bibliographystyle{IEEEtran}
\bibliography{refs}

\end{document}